\documentclass{aa}
\usepackage{graphics,epsfig,amsmath,amssymb,amstext}

\def\gsim{\lower.5ex\hbox{$\; \buildrel > \over \sim \;$}}
\begin{document}

    \title{High resolution $\gamma$-ray spectroscopy of flares on the east and 
west limbs of the Sun}

\author{M.J. Harris\inst{1} \and V. Tatischeff\inst{1} 
\and J. Kiener\inst{1} \and M. Gros\inst{2} \and G. Weidenspointner\inst{3}}
\institute{CSNSM, IN2P3/Universit\'{e} Paris-Sud, 91405 Orsay Campus, France
\and DSM/DAPNIA/Service d'Astrophysique, CEA Saclay, 91191 Gif-sur-Yvette,
France \and CESR, B.P. No. 4346, 31028 Toulouse Cedex 4, France}

\date{Received date; accepted date}
\authorrunning{Harris et al.} \titlerunning{Gamma Line Spectroscopy of Solar
Limb Flares}

\abstract{A new generation of Ge-based high-resolution gamma-ray spectrometers has allowed accurate measurements to be made of the profiles, widths and energies of the gamma-ray lines emitted in the impulsive phases of solar flares.  
Here we present measurements in two flares of the energies of the de-excitation lines of $^{12}$C and $^{16}$O at 4.4 and 6.1 MeV respectively by the Ge 
spectrometer SPI on board INTEGRAL, from which Doppler shifts are derived and 
compared with those expected from the recoil of $^{12}$C and $^{16}$O nuclei
which were excited by the impacts of flare-accelerated ions.  An anomalous 
Doppler measurement (in terms of recoil theory) has been reported by the Ge 
spectrometer {\em RHESSI\/} in a flare near the east limb, and explained by a 
tilt of the magnetic field lines at the footpoint of a magnetic loop away
from the vertical, and towards the observer.   This
might be interpreted to imply 
a significant difference between the Doppler shifts on the east and west limbs,
if it is a general phenomenon.
SPI observed both east and west limb flares and found no significant difference
in Doppler shifts.  We also measured the 
shapes and fluences of these lines, and their
fluence ratio to the 2.2 MeV line from the capture of flare-generated neutrons.
Analyses of both quantities using thick-target models parametrized by solar
physical and geometric quantities suggest that the two flares studied
here also have magnetic fields tilted towards the observer, though the
significance of the measurements is not high.}

\maketitle

\section{Introduction}

The paradigm for the origin of high-energy non-thermal radiations in solar
flares has for many years been that of the magnetic loop rising from the
solar surface in which a rapid release of energy (directly or indirectly
due to reconnection of the magnetic field configuration) accelerates ions
and electrons to high energies.  They are trapped in the loop and 
sooner or later encounter the dense material in the footpoints of the
loop (where it intersects the solar surface, probably at chromospheric
levels).  Here the ions may undergo nuclear reactions with the ambient 
material, generating secondary particles (notably neutrons) and $\gamma$-ray
lines from the de-excitation of the impacted nuclei and their fragments. 

Nuclear line characteristics --- amplitude, energy, width and shape --- 
contain a great deal of information about the accelerated beam and the
ambient gas, which was intensively studied in data from the {\em SMM\/}
and {\em CGRO\/}/OSSE missions 
(Share and Murphy 2001).  The advent in this century
of missions with finer energy resolution should in principle improve the
quality of this information, particularly that from line energies and shapes.
The dedicated solar mission {\em RHESSI\/} has presented measurements of
several flares (Shih et al. 2004)
and the spectrometer SPI on board INTEGRAL has also observed several
strong flares since its launch.
These both employ germanium detector technology
which achieves a FWHM resolution of 2--3 keV at energies around 1 MeV.

We report here SPI observations of two flares, with the focus being on the
measurements of $\gamma$-ray line energies and the Doppler shifts they
imply.  The reason for this focus may be understood in terms of an
idealisation of the magnetic loop paradigm described above: a single 
symmetrical loop which enters the surface vertically at each footpoint.
The beam travelling along the field lines may impact the target nuclei in
various directions, depending on the beam geometry (downward beam,
isotropic, downward hemisphere isotropic, fan beam etc.); 
the nuclei recoil in the directions of the impacts.
In general the lifetimes of the excited residual nuclear state are short
enough for de-excitation to occur in flight, the line being thereby
Doppler shifted.  As seen from Earth the shift is always red for a
downward beam; at the center
of the disk the component of the recoil velocity in the Earthward direction
is 100\% and the redshift is maximal, whereas on the limb the recoil is
transverse and the redshift zero.

By combining {\em SMM\/} data from many flares and binning them according to 
heliocentric angle, Share et al. (2002) obtained a correlation which
agreed with the expectations of the recoil model.  However the first flare
analyzed by {\em RHESSI\/} (2002 July 23, at a heliocentric angle 73$^{\circ}$)
had an anomalously large redshift for one so near the limb.  Smith et al.
(2003) gave a geometrical explanation in terms of the loop model: if the 
field lines do not emerge vertically from the surface but are directed towards 
Earth, the accelerated particles moving downwards along them impel the recoils 
away from Earth.

We shall discuss the measurements further in \S 4.  Here we note that, if
the phenomenon is $general$, then in the simple idealised picture there must 
exist cases where the field lines are directed away from Earth which give
rise to blue shifts.  If Earthward field line tilts 
at a given limb are a stable property of active regions
as such, then the maximal tilt away from Earth is seen at the opposite
limb.\footnote{ This picture only works if the loop is oriented in an east-west
({\em preceding-following\/}) direction.
A good analogy of a preceding-following asymmetry is
that of sunspot polarities: in an idealised picture, the preceding pole
(N or S according to hemisphere and sunspot cycle) is always closer to Earth
in regions coming into view round the east limb, but it disappears first in 
those going round the west limb, leaving
the following polarity (S or N) closer to Earth.}  We decided to test this
hypothesis by comparing the recoil Doppler shifts of nuclear lines in flares
on both limbs of the Sun.

Measurements of other nuclear line properties are of less interest, since
the two flares under consideration had much lower count rates (in SPI)
than many previously observed flares (the difficulty of observing solar
flares with SPI is discussed in the next section).  However by using
sophisticated thick-target models of the interactions at the footpoints,
it is possible to show that the same tilting of magnetic field lines
may have occurred in the two limb flares observed by SPI, to a lesser
degree than in the 2002 July 23 event (\S \S 4.2, 4.3).

\section{Observations and Analysis}

\subsection{Instrument and Observations}

The INTEGRAL payload was launched 17 October 2002 carrying two large 
$\gamma$-ray telescopes, IBIS and SPI; of these SPI is better equipped for
spectroscopy at $\gamma$-ray energies (see below).  The 3-day orbit is
highly eccentric and only 10\% of the time is spent in Earth's trapped
particle belt, with the instruments switched off.  For the 
remaining 90\% of the orbit SPI is in principle capable of acting as a
solar flare monitor.  Its normal operational mode is to point toward a
pre-selected target, and a number of constraints exist on where the
16$^{\circ} \times 16^{\circ}$ aperture \footnote {This is the fully-coded
inner section of the coded mask which SPI uses for imaging} may be pointed.
The solar panels providing power (defined as the spacecraft $\pm y$ axis) 
must face the Sun within 40$^{\circ}$
(Jensen et al. 2003), and IBIS must be between SPI and the Sun to provide
shadowing (the direction of IBIS is defined as 
spacecraft $+z$ axis).  Therefore SPI can never observe the Sun 
directly; the Sun is typically to the rear of the instrument, somewhere
on the ($-x,+z$) plane, where $+x$ is the pointing direction.

\begin{figure}
\centering
\epsfig{file=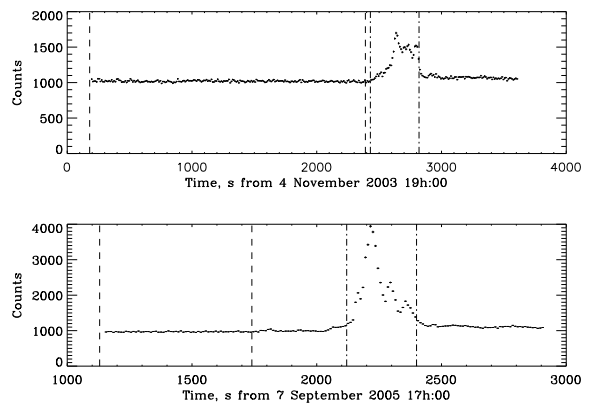,width=0.9\linewidth}
\caption{Count rate in the Ge detectors during the two flares.  Dashed
lines --- time intervals chosen for accumulation of the background before
the flare.  Dot-dashed lines --- time intervals of the flares.}
\end{figure}

The instrument itself consists of 19 hexagonal Ge detectors behind a
tungsten coded mask.  This ensemble is surrounded by a massive BGO
detector which defines the full aperture by operating in anticoincidence
to screen out photons from the side and rear.  The imaging capability
given by the mask is not used in the present work, since solar flare
photons always enter from the rear.  The 5 cm thick BGO shield 
is an efficient absorber of $\gamma$-rays from all side
and rear directions, including that of the Sun.  This represents a 
considerable loss of efficiency to the Ge detectors when they are
used to detect solar flare photons.  Other potential absorbers which
may intervene between the detectors and the Sun are the payload module
platform, occupying a large solid angle around the $-x$ axis, and the
IBIS detector plane, especially the PICsIT CsI layer. 

The Ge detectors have energy resolution $\sim 3.6$ keV at the energies
around 5000 keV in which we are interested.  Their number has been
reduced to 17 effectives by two failures, in 2003 December and 2004
July.  The resolution degrades
over time due to cosmic ray impacts, and is restored by an annealing
procedure at roughly 6-month intervals.

The two limb flares which we selected for analysis occurred on 2003 
4 November (S19W83, GOES class X26) and on 2005 7 September (S06E89,
GOES class X17).  The former was spawned by the same active region
as the X17 flare of 2003 28 October, results from which have been
published already (Kiener et al. 2006); by 4 November it had moved
from near the centre of the disk to the west limb.  The aspect of
the Sun with respect to SPI was very similar to that on 28 October,
being in the ($x,z$) plane 128$^{\circ}$ from the $+x$ axis (rather
than 122$^{\circ}$, Gros et al. 2004).  During the 2005 September flare
the Sun was close to the $z$-axis.  The $\gamma$-ray light curves of
the two flares are shown in Figure 1.  For the purpose of background
subtraction (see next section) we defined a period of flare activity
in each case, and also a preceding period of background when the Sun
was assumed to be quiescent.  These periods are also shown in Fig. 1.
Constraints on the choice of these periods are largely due to the
need to avoid periods of flare-associated charged particle bombardment,
as measured by the on-board INTEGRAL Radiation Environment Monitor
(Hajdas et al. 2003).  

\begin{figure}
\centering
\epsfig{file=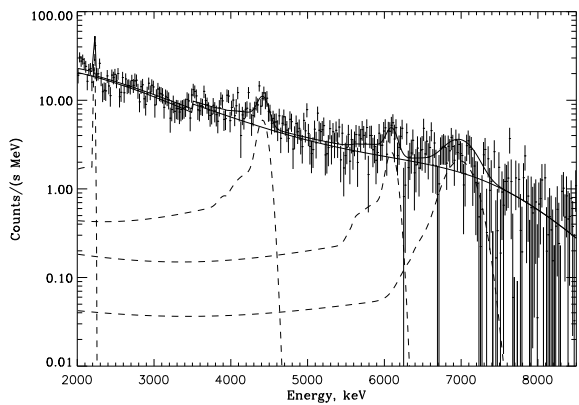,width=0.9\linewidth}
\caption{Background-subtracted spectrum of the 2003 November 4 flare.
Dashed lines --- $\gamma$-ray lines from neutron capture and $^{12}$C and
$^{16}$O de-excitation.  Full lines --- (lower)
continuum from underlying energy-dependent power law and nuclear
continuum; and (upper) total line-plus-continuum spectrum.}
\end{figure}

\subsection{Analysis}

From event-by-event detector-by-detector data files, we accumulated
spectra of the background and flare periods in the two cases over time
intervals of 10 s.  The event types included were single-detector
events, multiple-detector events with multiplicity 
up to 5, and events identified by the SPI
pulse-shape discrimination system (Vedrenne et al. 2003).  The events
were binned into energy intervals of either 1 keV or 20 keV.  These
spectra were corrected for dead time and calibrated in energy by
fitting Gaussian profiles to instrumental lines at well-known energies;
the lines were visible in the flare spectra as well as the preceding
background period, and we checked in the flare spectra that the
calibration was unchanged, although it was less accurately determined
than in the background phase.

The background subtraction was done by the same method as Kiener et al.
(2006), namely by measuring the ratios of the intensities of a sample of 
instrumental lines of varying half-lives during the background periods
to those in the flaring periods.  The background-subtracted flare spectra
at 20 keV resolution are shown in Figures 2 and 3.  
They were fitted between 2--10 MeV by spectral models
consisting of two continuum components and five lines at 2223, 4438,
6129, 6916 and 7115 keV, each of which consisted of three (or more)
components:\\

\begin{figure}
\centering
\epsfig{file=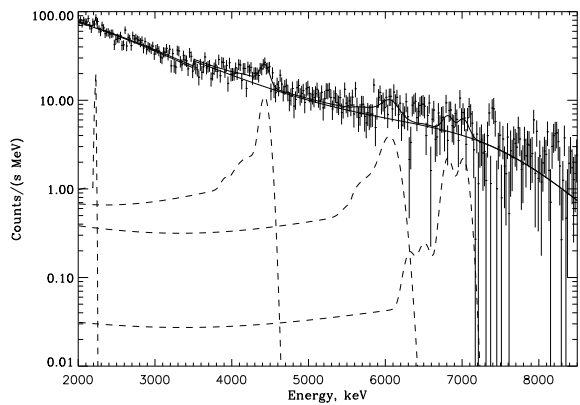,width=0.9\linewidth}
\caption{Background-subtracted spectrum of the 2005 September 7 flare.
The full and dashed lines are as in Fig. 2.}
\end{figure}

(a)  The continuum model consisted of analytic approximations to the
known solar flare continuum components: a steep underlying, basically 
electron-bremsstrahlung spectrum and a rather flat component at 
2--7.5 MeV due to nuclear reactions between accelerated
heavy ions and protons at rest.  We approximated the electron
bremsstrahlung component and the heavy-ion impact component below 3.5
MeV by a power law with a linearly energy-dependent index.  At 3.5
MeV the heavy-ion component increases sharply due to the onset of
accelerated carbon-plus-proton reactions.  We treated this as a 
second continuum component, constant up to 7.5 MeV where it was 
curtailed by a Woods-Saxon term.  Our motivation for this parametric
approach was that, since we will obtain physical information only
from lines, it becomes much easier to determine systematic errors
upon the lines from the continuum if the latter consists of
parameters which can be varied in a simple way.\\
(b)  The response of SPI to incident solar $\gamma$-ray lines was
modelled for an incident angle 122$^{\circ}$ from the $+x$ axis
in the ($x,z$) plane (see \S 2.1), using the Monte Carlo
simulation code MGGPOD (Weidenspointner et al. 2005) as described by Kiener 
et al. (2006).  The lines undergo Compton scattering, producing a
long tail at energies below the photopeak, and pair production followed
by single or double escape of the annihilation photons, producing
discrete lines shifted downward in energy by 511 or 1022 keV.  These
components were all included in our spectral model for the lines.  

The solar flare secondary neutrons producing the 2223 keV line are
not expected to undergo energetic impacts giving rise to recoil
Doppler shifts, but rather to thermalize before being radiatively
captured by protons almost at rest.  We included it in our analysis
mainly as a check on our calibration, using the spectrum at 1 keV
resolution.  Thermalization occurs in
relatively deep layers of the solar atmosphere, so that an additional
Compton tail component arising during these photons' escape from 
the Sun was calculated (Kiener et al. 2006).  The two lines around
7 MeV were found to be rather weak (see \S 3), but we included them
in our analysis because they may influence the energy and width of the
stronger 6129 keV line.  We generally forced them to be in a fixed
amplitude ratio of 1:1 and fixed their relative positions.  We
attempted to measure several lines below 2 MeV whose intensities in
ratio to other lines are diagnostic of the spectrum of the incident
flare particles: these are de-excitations of $^{24}$Mg (1369 keV),
$^{20}$Ne (1634 keV) and $^{28}$Si (1779 keV).

\section{Results}

Our measurements of the energies and widths of the nuclear lines are
presented in Table 1.  The errors are dominated by the statistical 
errors due to the short durations and low fluences of the flares.
The closeness of the neutron capture line energy to its rest energy of 2223
MeV suggests that our energy calibration is valid.  We also considered
different forms of underlying continuum from the energy-dependent power
law (\S 2.1), different energy onset of the heavy-ion induced nuclear
continuum, and different treatments of the complex of oxygen de-excitation
lines around 7 MeV (fixed amplitudes in various ratios, or independent
fitting).  The effects of these changes in our analysis were much more
serious on the amplitudes of the 4.4 and 6.1 MeV lines than on their
widths and energies, in which we are interested.  

A weak systematic
feature is seen in Fig. 3 around 7.4 MeV which is perhaps due to
the imperfect subtraction of a background $^{70}$Ge neutron 
capture line.  
This effect is not seen in the many Ge background lines at lower energies
and the feature is much weaker than the lines in whose energies we are
interested.  The differences between the line shapes at $\simeq 7$ MeV
in Figs. 2 and 3 are due rather to the fact that the coupled lines in
the 7 MeV feature are fitted by narrower widths in the 2005 7 September flare.

The de-excitation lines of $^{24}$Mg, $^{20}$Ne and $^{28}$Si were only
marginally detected, if at all ($2 \sigma$ at most), and we have therefore
given their amplitudes as $3\sigma$ upper limits in Table 2, where we present 
our measured line amplitudes.

\begin{table*}
\caption{Table 1. Line energies and widths (all values in keV)}
 ~\\
\begin{tabular}{lcccc}
\hline
Line & Energy & & Width FWHM & \\
rest & Nov 4 2003 (W) & Sep 7 2005 (E) & Nov 4 2003 (W) & Sep 7 2005 (E) \\
\hline
2223.2 & 2224.2$\pm 0.3$ & 2223.5$\pm 0.8$ & 4.0$\pm 1.9$ & 3.7$\pm 1.8$ \\
4438 & 4423$^{-13}_{+15}$ & 4432$^{-11}_{+13}$ & 161$_{+40}^{-29}$ &
 132$^{-38}_{+46}$ \\
6128 & 6101$\pm 20$ & 6076$\pm 34$ & 161$^{-48}_{+50}$ & 239$^{-32}_{+36}$ \\
6916 & 6894$\pm 22$ & 6817$\pm 31$ & 124$\pm 65$ & 155$\pm 50$ \\
\hline
\end{tabular}
\end{table*}

\begin{table*}
\caption{Table 2. Line amplitudes, counts s$^{-1}$}
~\\
\begin{tabular}{lcc}
\hline
Line energy, keV & 2003 November 4 & 2005 September 7 \\
        & $\theta = 85^{\circ}$ & $\theta = 89^{\circ}$ \\
\hline
1369 & $<$0.59 & $<$0.60 \\
1634 & $<$1.90 & $<$1.54 \\
1779 & $<$1.70 & $<$1.02 \\
2223 & 0.44 $\pm 0.09$ & 0.26 $\pm 0.12$ \\
4438 & 0.92$^{-0.18}_{+0.28}$ & 1.42$\pm 0.36$ \\
6128 & 0.39$^{-0.11}_{+0.16}$ & 0.83$^{-0.14}_{+0.18}$ \\
6916$^{a}$ & 0.20$\pm 0.10$ & 0.36$\pm 0.14$ \\
\hline
\end{tabular}

 ~\\
$^{a}$ The line at 7115 keV was fitted together with this one,\\
with amplitude fixed to be the same.\\

\end{table*}

\section{Discussion}

\subsection{Line energies and Doppler shifts}

The line energies measured in Table 1 can be compared with previous results,
and with the expectations of recoil theory, by dividing by the rest energy
and multiplying by the atomic mass number (Murphy \& Share 2006).  This is done for the 4438 keV
and 6128 keV lines in Fig. 4, which are compared as functions of cosine of
heliocentric angle with the measurements of
{\em SMM\/} (Share et al. 2002, 19 measurements in bins of heliocentric angle),
{\em RHESSI\/} (Share \& Murphy 2006, Murphy \& Share 2006),
and the point at heliocentric angle 25$^{\circ}$
measured earlier by SPI (Kiener et al. 2006).

The trend which would be expected in this diagram according to the recoil
model follows from a  very simple treatment of
the component of the recoil velocity in the
direction towards or away from Earth: $-V cos \theta$ at heliocentric
angle $\theta$ (for simplicity we assume a purely downward pencil
beam geometry).  The recoil velocity $V$ can be derived analytically 
from conservation of energy and momentum in such a very simple case
(non-relativistic, vertically-downward 
monoenergetic pencil beam in head-on collision):
$V \cong \frac{v}{1+\frac{M}{m}}$, where $v$ is the incoming proton velocity,
$m$ its mass and $M$ the ion's mass.  Share and co-workers (Share \& Murphy
1997; Share et al. 2003) have shown that an isotropic distribution in the
downward direction is to be preferred to a simple downward beam.  In this
case the naive treatment yields $V \cong \frac{\frac{1}{2}v}{1+\frac{M}{m}}$,
which is shown as the dashed line in Fig. 
4 for 10 MeV protons incident on $^{12}$C.

It is clear that the series of {\em SMM\/} measurements follow a trend
which is in quite good agreement with the dashed line.
The {\em RHESSI\/} measurements,
however, suggest a constant redshift as a function of $\theta$.  In
particular the 2002 July 23 {\em RHESSI\/} measurement (cos$\theta$=0.29)
has too large a redshift to 
be brought into agreement with the {\em SMM\/} trend nor even with our 
simplistic estimate (dashed line).

\begin{figure}
\centering
\epsfig{file=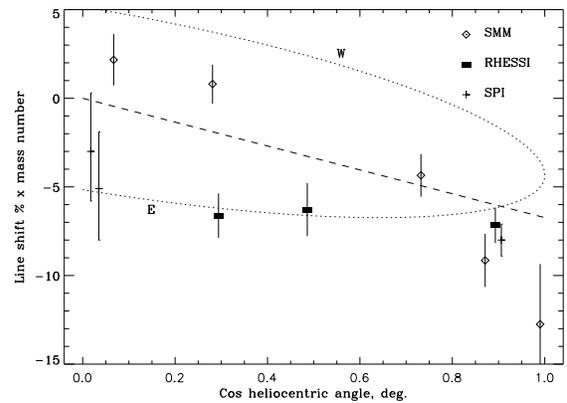,width=0.9\linewidth}
\caption{Doppler shifts of $^{12}$C and $^{16}$O
line energies from rest, multiplied by
atomic mass of recoiling nucleus and averaged.  Dashed line --- expected Doppler
shift from simple model of 10 MeV protons incident on ambient solar
atmosphere nuclei in a downward isotropic beam at 90$^{\circ}$.
Dotted lines --- expected shifts if the protons are incident
along magnetic field lines tilted at 50$^{\circ}$ to the horizontal.  If
this is the explanation for the {\em RHESSI\/} point at
cos$\theta$=0.29 towards the east limb, then the Doppler shifts should
follow the branches labelled "E" and "W" to the east and west of centre
respectively.  Sources of data points in text, \S 4.1.}
\end{figure}

Smith et al. (2003) therefore suggested that the accelerated ions were
beamed in a downward direction in this flare, but the magnetic field was
tilted by an angle $\sim 50^{\circ}$ to the surface, instead of $90^{\circ}$
as usually assumed.  In a study of $\alpha$-induced reactions in this
same flare Share et al. (2003) found evidence of the tilting, but ruled out
the downward beaming in favour of the more typical 
isotropic distribution in the downward hemisphere.

In terms of the analytic approach above, when the
magnetic field is tilted the redshift $-V cos \theta$ is
replaced by $-V sin (\theta + \phi)$  where $\phi$ is the angle between the
field at the footpoint and the solar surface.  We wish to test
the hypothesis that such tilts are a general and characteristic feature of 
magnetic loops in solar active regions.  The equation given here can then
be applied at any $\theta$, and it is obvious that it is double-valued
corresponding to values of $\theta$ to the east and west of the central
meridian.  The effect of this on our simple recoil-derived trend line
is to split it into two branches, which are shown by dotted lines in
Fig. 4 ($\phi = 50^{\circ}$ here).  Note that the east-limb branch now
agrees well with the {\em RHESSI\/} data point at cos$\theta$=0.29, as
intended.

The difference between the two branches of the dotted line in Fig. 4
is greatest at the limbs, where the west limb ought to exhibit a substantial
blue shift.  Our measurements of the two limb flares show no evidence for
any difference between them; if anything the west limb flare has a 
marginally greater redshift.  It lies $\cong$3 standard deviations away from
the blueshift which the simplistic model predicts.  We conclude that tilted
magnetic loop structures, though they undoubtedly exist on the Sun, are
not general and characteristic features of flare sites.

Compared with the other measurements in Fig. 4, our limb flare
measurements lie in between the trends visible in the {\em SMM\/} and
{\em RHESSI\/} data.\footnote{ The third SPI measurement shown in Fig. 4
(Kiener et al. 2006), by contrast, is in good agreement with both.}
The SPI measurements are consistent with a redshift trend proportional
to cos$\theta$, as in the naive recoil model, but the slope would be
smaller than that shown by the {\em SMM\/} data.  As a result of the 
rather large error bars, they are also consistent with the suggestion
in the {\em RHESSI\/} data of a constant modest redshift as a function of
$\theta$.  

Finally, we note that the {\em SMM\/} data points in Fig. 4
represent the averages of several flares.  Going back to the original
flare selection by Share \& Murphy (1995), we find that the bin closest
to the limb (which has the largest blueshift) contains only west-limb flares;
there is a considerable scatter in the flare-by-flare measurements, which
also have large error bars; this may conceal a pattern whereby some
west limb flares follow the behaviour expected from tilted magnetic field lines
(dotted line, upper branch, Fig. 4), whereas others are better explained by
perpendicular field lines (dashed line, Fig. 4).

We shall return to the question of tilted magnetic field lines in the
following sections, using a much more sophisticated model.

\subsection{Line widths and shapes}

Despite the relative weakness of the $^{12}$C and $^{16}$O lines in
the two flares, we also attempted to analyse their shapes, from which
much information can be derived, in principle.  Unlike the Doppler
shifts, which we found to be understandable from a rather simple
treatment of recoil theory, these shapes depend on multiple parameters
of the incident beam, the ambient gas and the viewing geometry, which
are incorporated in a complex model.  We
selected four of these parameters for detailed treatment.  First, the
beam geometry, which we allowed to take one of four extreme values ---
a downward conical beam with a variable opening angle, an isotropic
distribution in the downward hemisphere, a fan beam (horizontal 
at the base of the loop), and a beam undergoing intense pitch angle
scattering (the $\lambda$ parameter of Hua, Ramaty \& Lingenfelter
1989 being set to
30).  Second and third, the $\alpha$-to-proton ratio and power law
spectral index $s$ of the beam.  Fourth, except for the downward conical
beam, we considered the viewing
geometry in terms of an "effective" heliocentric angle $\theta$.  The true
heliocentric angle of these flares is of course known to be
very close to 90$^{\circ}$; however, were we to find our data
better fitted by a different value, this would most easily be explained
by a tilting of the field lines at the base of the loop, which is
the same hypothesis which we are testing with our recoil Doppler
measurements\footnote{ Note that in the following section we use a
different definition of effective heliocentric angle}.

\begin{figure}
\centering
\epsfig{file=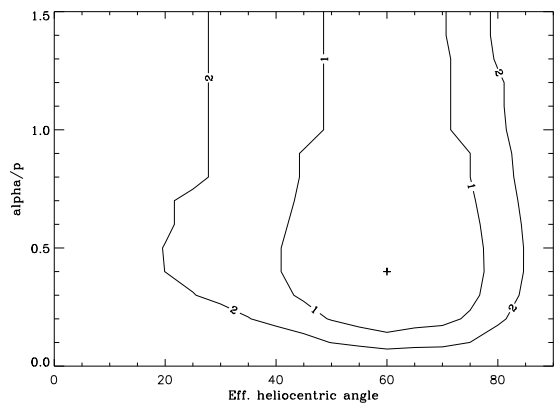,width=0.9\linewidth}
\caption{Best values of $\alpha$/$p$ and effective heliocentric angle
for the 2003 November 4 flare (cross), with the 1$\sigma$ and 2$\sigma$
contours around this point.  A downward isotropic beam is assumed;
best-fitting particle power law index $s = 3.5$.}
\end{figure}

For the three latter parameters we set up a grid of values, calculated
the line shapes at each point 
from the thick-target model described by Kiener et al. (2001),
and compared them with the data in Figs. 2 and 3 until the best
value of $\chi^{2}$ was found.  The grid values were restricted to
$0.05 \le \alpha$/$p \le 1.5$, $5^{\circ} \le \theta \le 85^{\circ}$,
and $2.0 \le s \le 5.0$.  Other input parameters were fixed; the most
important for the line fits, the C/O ratio, was found to give the best
fits if set at twice the standard solar value.  The lines and
continuum were fitted separately, the lines by the predicted shapes from
the thick-target model after fitting the continuum between 3.5--7 MeV,
where a power law approximation is sufficient.

As expected, only the weakest constraints result from this procedure, with
one interesting exception.  We illustrate the results for the $\alpha$/$p$
ratios in the two flares in Figures 5 and 6 where the 1$\sigma$ and
2$\sigma$ contours of $\chi^{2}$ are illustrated.  In the 2003 November 4
flare $\alpha$/$p$ is constrained only to be $>0.07$ at the 2$\sigma$ level
and in the 2005 September 7 flare to be $<0.9$ at this
level.  Similarly we found the particle power law index could only be 
constrained to $s > 2.0$ in both flares.

However Figs. 5 and 6 indicate that for both flares the effective heliocentric 
angle $\theta$ is better characterised, though the error bars are large.
Our formal results are $\theta = 60^{-19}_{+17}$ for the 
2003 November 4 flare ($s = 3.5$), and 
$\theta = 75^{-18}_{+19}$ for the 2005 September
7 flare, for $s = 4.0$.  The error bars are not linear, but it is
clear from Figs. 5 and 6 that neither of these is compatible with the known
heliocentric angle $\sim 90^{\circ}$ --- for the 2003 November 4 flare at about
the 2$\sigma$ level, and for the 2005 September 7 flare at the 1$\sigma$
level --- implying that the magnetic field lines in these flare footpoints
may be tilted.

The angles of tilt of these two limb flares (60$^{\circ}$--75$^{\circ}$
to the horizontal) are more modest than the value 50$^{\circ}$ 
proposed by Smith et al. (2003, see \S 4.1; 90$^{\circ}$ being the 
untilted value).  They are well within the error bars of the two points
for these flares in Fig. 4 --- if the tilt is towards the observer, given
the Doppler shift.
Thus our direct measurements of the tilt of the magnetic field lines
support its existence at the 1$\sigma$--2$\sigma$ level, 
but lead to the rather uncomfortable conclusion 
that all three of the tilts so far measured (two here and Smith et al. 2003)
are in the direction towards us.  No blue shift such as would be
expected from this scenario has yet been seen.

\begin{figure}
\centering
\epsfig{file=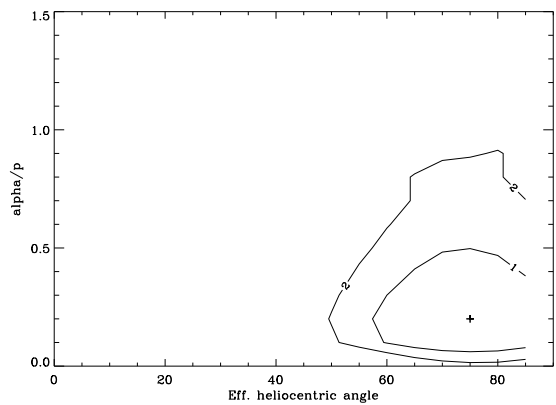,width=0.9\linewidth}
\caption{Best values of $\alpha$/$p$ and effective heliocentric angle
for the 2005 September 7 flare (cross), with the 1$\sigma$ and 2$\sigma$
contours around this point.  A downward isotropic beam is assumed;
best-fitting particle power law index $s = 4.0$.}
\end{figure}

\subsection{Line fluence ratios}

The Doppler shifts of the nuclear de-excitation lines can be understood
by very simple recoil arguments from the standard magnetic loop paradigm,
as we saw above.  Similarly simple arguments can explain not only the lack of
Doppler shift of the neutron capture line at 2223 keV (\S 2.2) but also its
amplitude.  The key point here is that neutron capture cross sections
increase as the neutrons slow down, so that they traverse a considerable
thickness of matter before being captured by the ambient hydrogen.  As
noted in \S 2.2, the resulting photons are Compton scattered (effectively
absorbed) on their way out to the surface.  The path length to the
surface is proportional to sec$\theta$ by the usual limb-darkening
arguments, so the absorption ought to be very severe for limb flares.
This is borne out by the low fluences measured for this line in Table
2, and by their trend with $\theta$.

More information is obtained from applying detailed models of the kind
discussed above (\S 4.2) to the ratio of the fluences in the 2223 keV
line to the sum of the $^{12}$C and $^{16}$O de-excitation lines.  These
fluences are obtained from the line amplitudes in Table 2 by first
dividing by the effective areas which Kiener et al. (2006) found for the
2003 October 28 flare at 2223, 4438 and 6129 keV to yield fluxes.  If
these fluxes are to be multiplied by the flare durations in Fig. 1,
account must be taken of the fact that the neutron capture line
is delayed relative to the others by the thermalization time
necessary for the neutrons to slow down to be captured.  The 2223
keV line is therefore underestimated if some neutrons continue to
be captured in a "tail" beyond the right-hand dot-dash line in Figs.
1a and 1b.  This late emission is too weak to be measured directly in
our two flares, given the low count rates in Table 2.  It was 
investigated in the 2003 October 28 flare by Kiener et al. (2006), who
derived a conservative correction factor of +10\%$\pm$10\% to be
applied to the 2223 keV line fluence.  When the fluxes are multiplied
by the flare durations after applying this correction, the fluence ratios
(2223 keV)/(4438 keV + 6129 keV) are found to be $0.33^{-0.11}_{+0.09}$
for the 2003 November 4 flare, and $0.11 \pm 0.06$ for the 2005
September 7 flare.

These ratios measured by SPI are shown in Fig. 7, together with the ratios
measured with SMM for nine flares at heliocentric angles less than 80$^{\circ}$
(Share \& Murphy 1995) and the ratio measured with {\it CGRO}/OSSE for the
1991 June 4 flare (Murphy et al. 1997).
These data are compared with theoretical
predictions for various parameters of the accelerated ion beam: the power law
spectral index, the $\alpha$/$p$ ratio and the angular distribution of
the accelerated particles. For the production of the 2223 keV line, we used
the Monte Carlo simulation code developed by Hua et al. (2002). This code
can calculate the depth, energy, and angular dependences of the
production of neutrons and their subsequent capture $\gamma$-ray 
line emission in
the solar atmosphere as a function of the parameters of the accelerated
particle beam, and then calculate the attenuation of the escaping 2223 keV 
photons as a function of the flare heliocentric position. We used the solar
atmosphere model of Avrett (1981). We have checked with the models of
Vernazza
et al. (1981) and Beebe et al. (1982) that the calculated ratios do not
significantly depend on the assumed density profile in the chromosphere and
photosphere.

The 4.44 and 6.13 MeV deexcitation lines from ambient $^{12}$C and $^{16}$O,
respectively, are produced in reactions of energetic protons and
$\alpha$-particles with ambient $^{12}$C, $^{14}$N, $^{16}$O and $^{20}$Ne.
We calculated the production of these two lines from the cross sections of
Kozlovsky et al. (2002), within the thick target interaction model of
Ramaty (1986). We
took into account the line attenuation in the solar atmosphere from the
calculations of Hua et al. (1989). These authors have shown that the
attenuation of the 4.44 and 6.13 MeV lines is significant only if the flare
is very close to or behind the limb of the Sun
($\theta$$\gsim$88$^{\circ}$).
The de-excitation gamma-ray lines are much less attenuated than the 2223 keV
neutron capture line, because they are produced higher in the solar
atmosphere.

The measured fluence ratios shown in Fig. 7 decrease with increasing
$\theta$ (i.e. decreasing cos($\theta$)) as a result of the 2223
keV line attenuation discussed at the
beginning of the section. We see in Fig. 7a that the theoretical curves for
$s$=3.5 and $\alpha$/$p$=0.1, and for $s$=4 and $\alpha$/$p$=0.5
provide a rather good interpretation of the data for most of the flares. It
is remarkable that the measured heliocentric angle dependence of the fluence
ratios can be satisfactorily reproduced with the same set of ion beam
parameters. However, the data for both the 1982 December 7 flare at
$\theta$=80$^{\circ}$ and the 2003 November 4 flare
($\theta$=85$^{\circ}$) fall significantly below these three curves. It is
possible that the accelerated particle spectrum was softer in these two
flares than in most of the other flares (see Fig. 7a). But it is also
possible
that the relatively low fluence ratios of these two events are due to an
inclination of the magnetic loops that would lead to an increase of the
2223 keV line attenuation.

\begin{figure}
\centering
\epsfig{file=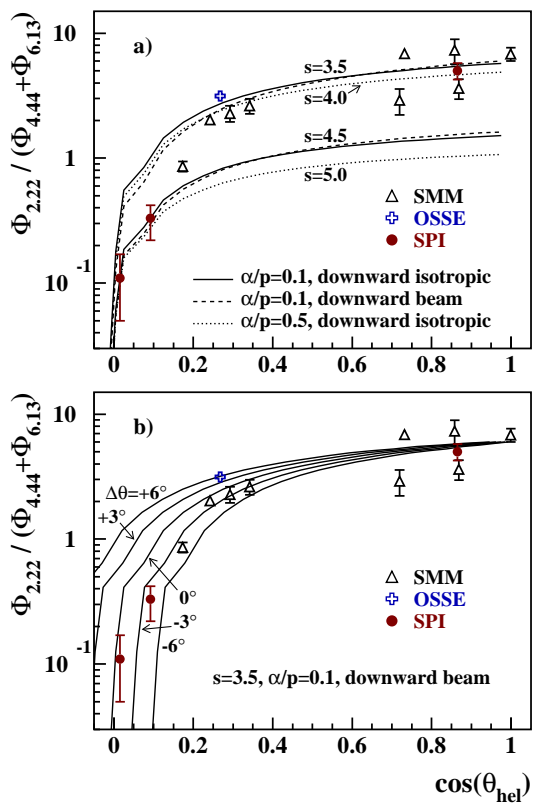,width=0.9\linewidth}
\caption{Comparison of calculated and measured ratios of the 2.223 MeV
neutron-capture line fluence to the summed 4.44 and 6.13 MeV $^{12}$C and
$^{16}$O line fluences, as a function of cosine of heliocentric angle.
SMM: Share and Murphy (1995); OSSE: Murphy et al. (1997); SPI: Kiener et al.
(2006) and this work. (a) Calculated ratios for various accelerated ion
compositions, angular distributions and energy spectra. Solid curves:
$\alpha$/p=0.1, downward isotropic distribution, $s$=3.5 and 4.5; dashed
curves: $\alpha$/p=0.1, pencil beam downward distribution, $s$=3.5 and 4.5;
dotted curves: $\alpha$/p=0.5, downward isotropic distribution, $s$=4 and 5.
(b) Calculated ratios for shifts in heliocentric angle of 0, $\pm$3$^{\circ}$
and $\pm$6$^{\circ}$ to simulate an inclination of the magnetic field
lines at the
base of the loop (see text). Here the calculations are for $s$=3.5,
$\alpha$/p=0.1 and a pencil beam downward distribution.}
\end{figure}

To simulate the expected effect of an inclination of the loop magnetic field
on the 2223 keV line emission, we simply replaced in Fig. 7b the true
$\theta$ by $\theta + \Delta \theta$, in such a way that positive
(resp. negative) values of $\Delta \theta$ would correspond to a tilt of the
magnetic loop away (resp. toward) the observer. The results of this figure
suggest that the two limb flares observed with SPI could be tilted toward us
and that the angle of the tilt for the 2003 November 4 flare could be more
important that the one for the 2005 September 7 flare, which is in
qualitative
agreement with the conclusions of the line shape analysis (\S 4.2).

The "effective" heliocentric angle used here has
a different meaning than that previously used in the line shape analysis.
Here it corresponds to the position of a representative flare with a
magnetic
field perpendicular to the solar surface that would give about the same
space
distribution of the neutron capture line emission in the solar atmosphere
as the flare at $\theta$ with inclined magnetic field lines. Detailed
simulations of 2223 keV line emission from solar flare tilted magnetic loops
are required to assess the validity of this approach and to link the
parameter
$\Delta \theta$ to the inclination of the loop.

\section{Conclusions}

The signal-to-noise ratios in the two flares discussed in this paper were
rather weak, so rather little information can be deduced concerning most
of the flare acceleration and transport physics.  However, two
conclusions can be drawn concerning the possibility that, in contradiction
to the simple picture in which flares particles are accelerated along
magnetic field lines arranged in a loop structure entering the solar
surface vertically, the angle of the field lines is in fact tilted to the
vertical.

First, anomalous Doppler shifts of $\gamma$-ray line energies (red if
the tilt is towards us, blue if away) are a disgnostic of this effect.
If the tilt is a general and characteristic property of magnetic fields
in flaring active regions, one would expect, not only to see Doppler
blueshifts of the lines, but that (since the
effect is greatest at the solar limb) there is a significant difference
between redshifted lines on one limb and blueshifts on the other ---
since the best $\gamma$-ray evidence for the effect is an anomalous 
redshift seen by {\em RHESSI\/} in a flare rather close to the
east limb flare (Smith et al. 2003), this would mean
significant blueshifts in lines from west-limb flares.  We found no
difference between the Doppler shifts in lines from flares from east
and west limbs (Fig. 4).

Second, measured line shapes and fluence ratios are also functions of the
tilt of the magnetic field lines, and we have searched the data for our
two flares for evidence that they themselves exhibited tilts.  We
find weak evidence in both cases for a tilt of the magnetic field
lines towards the observer (although smaller than the tilt found by
Smith et al. (2003) in July 2002). 

The rather unexpected overall conclusion is that of the flares with
$\gamma$-ray measurements of tilted magnetic field lines, in all three
cases the tilt was towards the observer, and in no case away.  It must
be acknowledged that the significance of two of these measurements
is low (\S 4.2).  Nevertheless, it is clearly important to obtain
$\gamma$-ray measurements of a larger sample of flares in order to
find the predicted blueshifted lines and other indications of magnetic field
lines tilted away from the observer.
 
\acknowledgements{This work was supported by the Universit\'{e} de Paris
XI (Paris-Sud).  Based on observations with the INTEGRAL satellite, an 
ESA project
funded by the member states (especially the PI countries: Denmark,
France, Germany, Italy, Switzerland, and Spain), the Czech Republic and
Poland, with the participation of Russia and the USA.}

\end{document}